\pdfoutput=1

\documentclass[11pt]{article}

\usepackage[]{EMNLP2023}

\usepackage{graphicx}
\usepackage[export]{adjustbox}
\usepackage{bm}
\usepackage{amssymb}
\usepackage{float}
\usepackage{booktabs}
\usepackage{multirow}
\usepackage{subfigure}
\usepackage{makecell}
\usepackage{amsmath} 
\usepackage{url}
\usepackage{times}
\usepackage{latexsym}
\usepackage{colortbl}

\usepackage[T1]{fontenc}

\usepackage[utf8]{inputenc}

\usepackage{microtype}

\usepackage{inconsolata}

\title{VKIE: The Application of Key Information Extraction on Video Text}

\author{Siyu An$^1$\footnotemark[1], Ye Liu$^2$\footnotemark[1], Haoyuan Peng$^3$ and Di Yin$^1$ \\
  $^1$Tencent YoutuLab $^2$Nvidia $^3$Learnable.ai \\
    \texttt{\{siyuan, endymecyyin\}@tencent.com} \\
    \texttt{liuyebug@126.com}, \texttt{haoyuan.peng@learnable.ai}}

\begin{document}
\maketitle
\begin{abstract}




Extracting structured information from videos is critical for numerous downstream applications in the industry.
In this paper, we define a significant task of extracting hierarchical key information from visual texts on videos. 
To fulfill this task, we decouple it into four subtasks and introduce two implementation solutions called  PipVKIE and UniVKIE. PipVKIE sequentially completes the four subtasks in continuous stages,
while UniVKIE is improved by unifying all the subtasks into one backbone.
Both PipVKIE and UniVKIE leverage multimodal information from vision, text, and coordinates for feature representation. 
Extensive experiments on one well-defined dataset demonstrate that our solutions can achieve remarkable performance and efficient inference speed. 


\end{abstract}

\renewcommand{\thefootnote}{\fnsymbol{footnote}} 
\footnotetext[1]{These authors contributed equally to this work.}

\section{Introduction}

Extracting information from video text is an essential task for many industrial video applications, i.e., video retrieval\citep{radha2016video}, video recommendation\cite{yang2007online}, video indexing\cite{yang2011lecture}, etc. Visual text embedded in videos usually carries rich semantic descriptions about the video contents, and this information gives a high-level index for content-based video indexing and browsing.  

Conventional methods utilize OCR \cite{liao2018rotation,tian2016detecting,zhou2017east} to extract visual texts from videos frames and employ text classification techniques \cite{le2018convolutional,li2020improving} to categorize the extracted content.
However, these methods suffer from two significant shortcomings:
1) Visual texts are typically coarse-grained at the segment level, and are unable to capture fine-grained information at the entity level, which is critical for downstream tasks.  
2) Traditional methods have not fully utilized the fusion of features from different modalities.

\begin{figure}[]
        \centering
            \includegraphics[height=7cm]{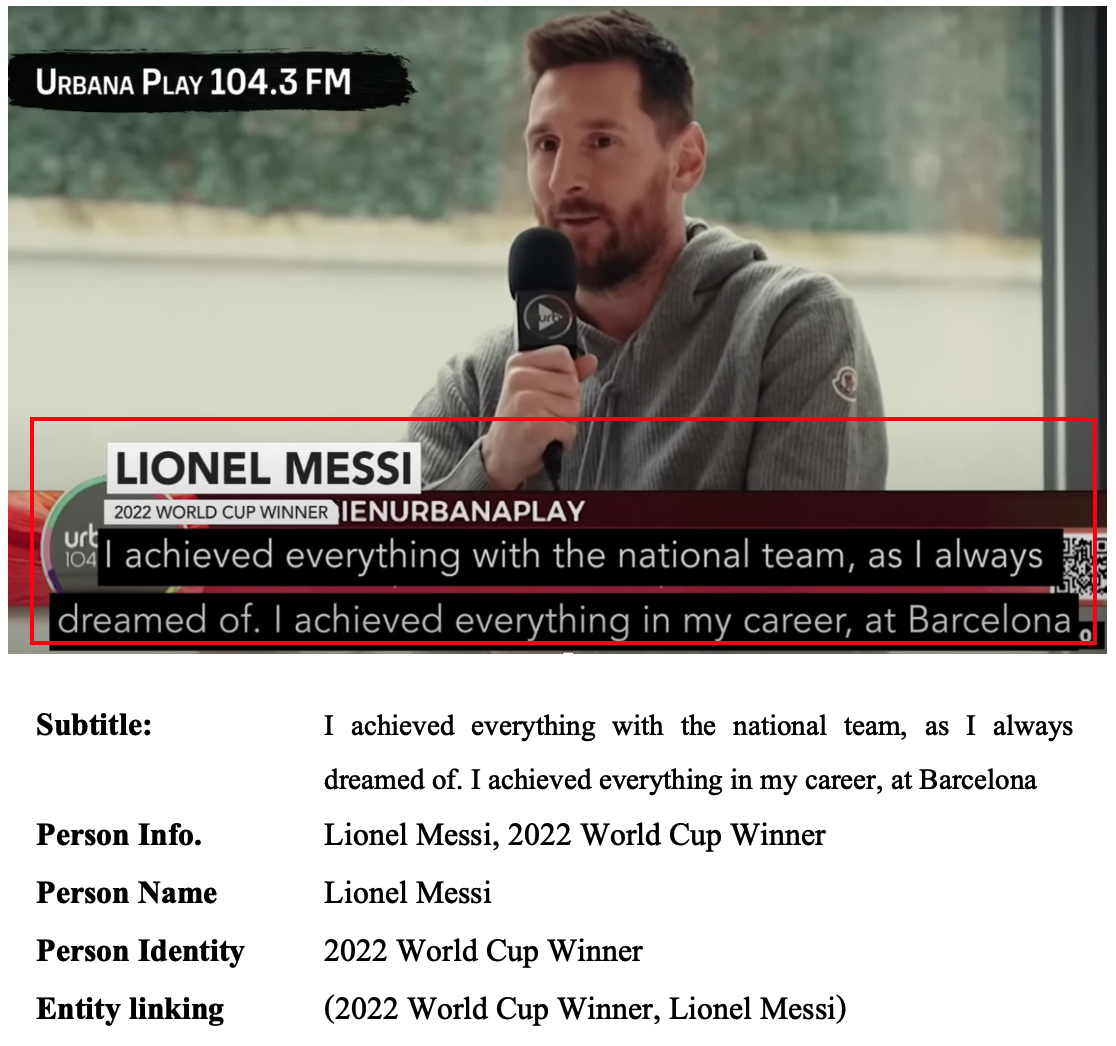}
        \caption{An example of hierarchical key information extracted by VKIE in a video frame \cite{CGTN_2023}.}
        \label{fig:fig1}
        \vspace{-2ex}
\end{figure}

Therefore, in our work, we introduce a novel industrial task for extracting key information from video text and exploring the relationship between entities, which we refer to as VKIE.
The task aims to extract valuable hierarchical information from visual texts, explore their relationships, and organize them in structured forms. 
This approach enables effective management and organization of videos through the use of rich hierarchical tags, which can be utilized to index, organize, and search videos at different levels. Figure \ref{fig:fig1} provides an example of the hierarchical key information extracted by VKIE, where subtitles are captured at the segment level, and personal information is organized with names and identities at the entity level.

To enhance clarity, we decompose VKIE into four subtasks: text detection and recognition (TDR), box text classification (BTC), entity recognition (ER), and entity linking (EL). While the first subtask, TDR, is typically accomplished using off-the-shelf OCR tools, our work concentrates on the remaining three subtasks of BTC, ER, and EL.

Since TDR outputs all boxes with text content and coordinates information, there are massive useless texts, such as scrolling texts and blurred background texts, which could have side effects on downstream tasks. BTC aims to eliminate these useless texts and find valuable categories, such as title, subtitle, and personal information.




Although the BTC method can obtain segment level information, the results are relatively coarse-grained and will limit its deployment to many downstream applications. 
For example, in video-text retrieval, the query is usually in different forms, such as  keywords, phrases, or sentences. In video indexing, a video is required to be stored with hierarchical tags. To address these issues, we designed ER to extract entities from text segments and EL to explore the relations among the entities. With this structured information, videos can be well managed with rich hierarchical information at the entity and segment levels.

In this paper, we present two solutions that have been deployed in our industry system. The first approach, called PipVKIE, involves performing the tasks sequentially, which serves as our baseline method. 
The second approach, called UniVKIE, achieves better performance and efficiency by more effectively integrating multimodal features.


In summary, our contributions are as follows:

(1) We define a new task in the industry to extract key information from video texts. By this means, structured information could be effectively extracted and well managed at hierarchical levels.

(2) We introduce and compare two deployed solutions based on the framework includes TDR, BTC, ER, and EL. Experiments show our solutions can achieve remarkable performance and efficient inference speed.

(3) To make up the lack of datasets, we construct a well-defined dataset to provide comprehensive evaluations and promote this industrial task.

\section{Approaches}



\begin{figure*}[h]
    \centering
    \includegraphics[width=1\linewidth]{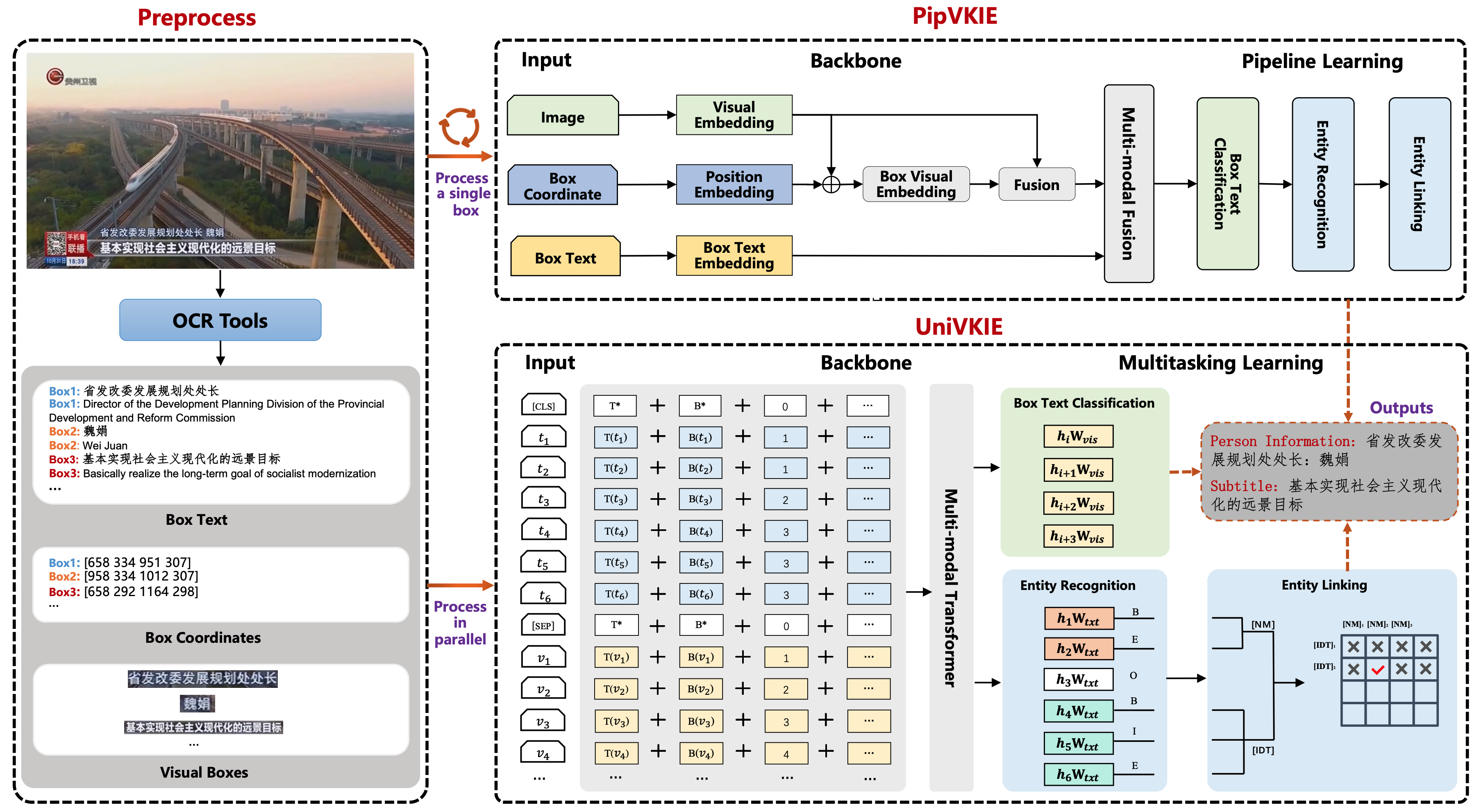}
    \centering
    \caption{The overall architecture of two deployed solutions PipVKIE and UniVKIE.}
    \label{fig:structure}
\end{figure*}

\subsection{PipVKIE}

The PipVKIE solution fulfills three subtasks of BTC, ER, and EL in a sequential pipeline and processes a single visual box at a time. In this process, BTC acts as a filter, selecting only the valuable text segments. After BTC performs, ER is carried out only on the segments selected by BTC. Similarly, when performing EL, only the entities extracted by ER are inputted, while other irrelevant information is filtered out.

\paragraph{BTC}

In our design, the objective of BTC is to categorize the text segments that appear on the OCR boxes into different classes, such as titles and subtitles. As illustrated in Fig.\ref{fig:structure}, in PipVKIE, BTC takes the visual and textual features as input and outputs the corresponding class label.
Specifically, for visual modality, in contrast to conventional approaches that usually use the classical VGG \cite{simonyan2014very} or ResNet-based \cite{he2016deep} network, we construct a shallow neural network as the backbone. In fact, we observe that texts differ in low-level features of colors and fonts, thus the above-mentioned deeper networks are abandoned as high-level semantic information is extracted. Consequently, transformer \cite{vaswani2017attention} is selected as the backbone of textual extraction. 

The fusion of multimodal features is a critical step in obtaining the multimodal representation of one box. The process of visual and positional modalities is shown below:
\begin{equation}
\bm{h}_{vb} = {\rm Trans} ({\rm{ROIAlign}} (\bm{h}_{vf}, \bm{h}_{p}), \bm{h}_{vf})
\end{equation}
where $\bm{h}_{vf}$ is visual embedding of frame directly obtained by CNN \cite{krizhevsky2012imagenet}, and $\bm{h}_{p}$ is positional embedding of box obtained by coordinates respectively. 
Firstly, ROIAlign \cite{he2017mask} is utilized to extract visual box embedding conditioned on $\bm{h}_{p}$ and $\bm{h}_{vf}$. Then, we take the transformer \cite{vaswani2017attention} to learn the implicit relation between a box and its corresponding frame, which denoted as $\bm{h}_{vb}$. The visual box embedding $\bm{h}_{vb}$ and the textual box embedding $\bm{h}_{tb}$, which is obtained by applying the transformer encoder on text, are simply concatenated to obtain the final multimodal vector representation $\bm{h}_{b}$. Subsequently, we perform softmax classification by multiplying $\bm{h}_{b}$ with trainable weight parameters.

\paragraph{ER}
Contrary to commonly known NER in flat text \citep{lample2016neural}, the goal of ER in VKIE is to identify entities from a single video frame. In this context, factors such as the entity's position and background features can significantly influence the recognition process.
In PipVKIE, what we need to accomplish at this stage is the extraction of entities from the valuable text segments selected in the previous step.
We obtain the hidden representation of text tokens by transformer encoder, and then predict their tags with the BIO2 tagging
schema \cite{sang1999representing}. 



\paragraph{EL}
EL aims to explore the relations between the extracted entities in each frame. 
Specifically, let ${\bm{h}_{p}^\textrm{N}}$
denote the hidden representation respect to $p$-th entity of the category \textit{Name}, ${\bm{h}_{q}^\textrm{I}}$ denote the hidden representation respect to $q$-th entity of the category \textit{Identity}, the representation of each entity is generated by the average pooling of text tokens.
Subsequently, in each frame, we build the matrix ${D}$ as inputs for the classifier. The element of ${D}$ is described in Eq.\ref{eq:doclm_taskttt}, where ${D}(p,q)$ represents the vector concatenated with the hidden representations of the entity pair $[{\bm{h}_p^\mathrm{N}, \bm{h}_q^\mathrm{I}}]$. 
\begin{equation}
\begin{split}
\label{eq:doclm_taskttt}
D(p,q) = [{\bm{h}_{p}^\textrm{N}}, {\bm{h}_{q}^\textrm{I}}]
\end{split}
\end{equation}



\subsection{UniVKIE}
Although PipVKIE is effective in practice, we have identified several problems with it: 1) PipVKIE does not effectively utilize the layout relationships between different boxes within the same frame. 2) The three tasks (BTC, ER, and EL) are trained separately and cannot benefit from each other. 3) Processing only one box at one time during inference is not efficient enough. 
To tackle the challenges posed by PipVKIE, we propose UniVKIE, a unified model that processes all boxes of each frame in parallel. UniVKIE leverages a shared multimodal backbone and employs a multitask learning approach. Fig.\ref{fig:structure} provides an overview of our model's architecture.

\subsubsection{Multimodal Backbone}

Similar to the model structure defined \citep{li2021structext, xu2020layoutlm, xu2020layoutlmv2, hong2022bros}, we utilize a shared multimodal backbone for the three tasks.
Given a frame of video,  we firstly apply OCR to obtain text recognition results which could be described as a set of 2-tuples including $M$ text segments and box coordinates. Then, we concatenate these $M$ text segments from top left to bottom right into one text with length $N$.  In this concatenated text, let $v_i \in \{v_1, v_2, ...v_M\}$  denote the $i$-th visual token with respect to $i$-th box and $t_j \in \{t_1, t_2, ...t_N\}$ denote the $j$-th token of text. Then we add $\texttt{[CLS]}$, $\texttt{[SEP]}$ and pad the sequence to fixed length $L$. The input sequence is established as the format in {Eq.\ref{eq:structext}}.

\begin{equation}
\begin{split}
\label{eq:structext}
S = \{\texttt{[CLS]}, t_1, \dots ,t_N, \texttt{[SEP]},\\ v_1, \ldots ,v_M, \texttt{[PAD]}, \dots \}
\end{split}
\end{equation}
UniVKIE benefits from this structure in two aspects: 1) visual token and text token can interact with each other, thus the  feature representation is reinforced by multimodal fusion. 2) the relations between boxes are  explored to  fully extract layout information. 3) all boxes in  each frame  are processed in  parallel  in these concatenated form.

\subsubsection{Multitask Learning}
While the models corresponding to the three subtasks are trained separately in PipVKIE, UniVKIE unifies these subtasks and employs a multitask learning approach\citep{vandenhende2020revisiting} to jointly train the model. As illustrated in Fig.\ref{fig:structure},
UniVKIE takes the embeddings of $M$ text segments defined in Equation \ref{eq:structext} as input to the BTC branch, which outputs the categories of 
$M$ boxes. The ER branch takes the $N$ tokens in the text concatenated by all box texts as input to identify the entities, which are then passed to the EL branch to explore their relationships.

By summing the losses of the three subtasks, we calculate the final loss as follows:
\begin{equation}
\mathcal L = \alpha \mathcal L_{BTC} + \beta \mathcal L_{ER} + (1-\alpha-\beta) \mathcal L_{EL}
\label{eq:doclm_joint_loss}
\end{equation}
where $\mathcal L_{BTC}$, $\mathcal L_{ER}$, and $\mathcal L_{EL}$ is the loss of BTC, ER and EL respectly, $\alpha$ and $\beta$ are hyperparameters to make trade-offs. 

\section{Experiments}
\subsection{Experimental Setup}
\label{Experimental Setup}
\paragraph{Dataset} To promote the new task, we have created a real-world dataset consisting of 115 hours of videos collected from 88 different sources. In preprocess, we uniformly sampled 23,896 frames from these videos and obtained over 123k visual boxes with text segments and coordinates by an off-the-shelf OCR tool. Afterwards, the dataset was carefully annotated and strictly checked by 8 professional annotators. Further details about the dataset are shown in Table \ref{tab:dataset description}. 

\paragraph{Metrics and Implementation Details} We evaluate the performance of BTC, ER, and EL by Precision (P), Recall (R), F1-score, and Accuracy (Acc). To ensure the reliability of our results, we conducted ten runs with distinct random seeds for each setting and report the average results obtained from these runs. Details of the  hyperparameters settings for PipVKIE and UniVKIE are presented in Table \ref{tab:hyperparameters of PipVKIE} and Table \ref{tab:hyperparameters of UniVKIE} respectively.

\subsection{Experimental Results}
\begin{table*}[h] 
\centering
\scalebox{0.8}{
    \begin{tabular}{l|*3l|*3l|*3l|*3l|l} 
    \hline
    \rowcolor{green!20!blue!10} \multicolumn{14}{c}{\textbf{BTC Task}}\\
    \hline
    \multirow{2}*{\textbf{Methods}} & \multicolumn{3}{c|}{\textit{Title}} & \multicolumn{3}{c|}{\textit{Person Info}} & \multicolumn{3}{c|}{\textit{Subtitle}} & \multicolumn{3}{c|}{\textit{Misc}} & Avg \\
    \cline{2-14}
    {} & P & R & F1 & P & R & F1 & P & R & F1 & P & R & F1 & Acc\\ 
    \hline
    BERT & 86.32  & 83.58 & 84.93 & 92.30  & 87.46 & 89.81 & 91.63  & 88.21 & 89.89 & 85.35 & 82.84 & 84.08 & 86.00 \\
    xlm-RoBERTa & 89.17  & 86.91 & 88.02 & 92.85  & 89.95 & 91.38 & 83.19 & 85.31 & 84.24 & 85.97 & 89.00 & 87.46 & 87.87\\
    ResNet-50 & 73.43  & 66.56 & 69.84 & 84.51  & 79.38 & 81.86 & 79.24 & 78.51 & 78.87 & 85.80 & 76.99 & 81.16 & 77.57\\
    PipVKIE   & 95.10 & 92.19 & 93.62 & 95.58 & 89.48 & 92.43 & 95.28 & 91.17 & 93.18 & 95.71 & 98.45 & 97.06 & 95.57   \\
    UniVKIE & 84.37 & 86.42 & 85.38 & 98.90 & 98.77 & 98.83 & 90.36 & 98.74 & 94.36 & 99.53 & 97.25 & 98.37 & \textbf{97.22\dag} \\
    \hline
    \rowcolor{green!20!blue!10} \multicolumn{10}{c|}{\textbf{ER Task}} & \multicolumn{4}{c}{\textbf{EL Task}}\\
    \hline
    \multirow{2}*{\textbf{Methods}} & \multicolumn{3}{c|}{\textit{Name}} & \multicolumn{3}{c|}{\textit{Identity}} & \multicolumn{3}{c|}{Avg} & \multicolumn{2}{c|}{\multirow{2}*{\textbf{Methods}}} & \multicolumn{2}{c}{Avg}\\
    \cline{2-10}\cline{13-14}
    {} & P & R & F1 & P & R & F1 & P & R & F1 & \multicolumn{2}{c|}{} & \multicolumn{2}{c}{Acc}\\ 
    \hline
    PipVKIE$^{\ast}$    & 92.92 & 92.08 & 92.50 & 74.43 & 77.30 & 75.84 & 84.71 & 85.69 & 85.19 & \multicolumn{2}{c|}{PipVKIE$^{\ast}$} & \multicolumn{2}{c}{81.51}  \\
    PipVKIE & 92.78 & 88.45 & 90.56 & 73.99 & 75.46 & 74.72 & 84.32 & 82.82 & 83.56 & \multicolumn{2}{c|}{PipVKIE} & \multicolumn{2}{c}{69.33}  \\
    UniVKIE$^{\ast}$ & - & - & - & - & - & - & - & - & - & \multicolumn{2}{c|} {UniVKIE$^{\ast}$} & \multicolumn{2}{c}{79.96} \\
    UniVKIE & 97.39 & 97.81 & 97.60 & 90.38 & 91.30 & 90.84 & 94.26 & 94.91 & \textbf{94.58\dag} & \multicolumn{2}{c|}{UniVKIE} & \multicolumn{2}{c}{\textbf{71.61\dag}} \\
    \hline
    \end{tabular}
}
\caption{Experimental results of BTC, ER, and EL. 
* indicates the results obtained by replacing the prediction of the upstream task with ground truth. - indicates the meaningless results, since for UniVKIE, ER does not rely on BTC in the pipeline. † indicates that UniVKIE performs better with p-value < 0.05 based on paired t-test.
{\label{tab:Main Results}}}
\end{table*}

\subsubsection{BTC}
The upper part of Table \ref{tab:Main Results} presents the performance of BTC. To evaluate how modality contributes to performance, we also take unimodal methods for comparison. 
This includes two text backbones, BERT \cite{devlin2018bert} and xlm-RoBERTa \cite{conneau2019unsupervised}, as well as ResNet-50 \cite{he2016deep}, which serves as a visual backbone.
Our results show that PipVKIE and UniVKIE outperform unimodal methods, with UniVKIE performing better than PipVKIE. This demonstrates the superiority of utilizing multimodal information and the unifying strategy.

\subsubsection{ER}

In PipVKIE, subtasks are completed in sequential stages, which means that errors can accumulate in the downstream task ER after BTC. To isolate the accumulated error, we evaluated the performance of ER by replacing the prediction of BTC with the ground truth. The performance of ER is shown in the bottom left of Table \ref{tab:Main Results}, where $\rm{PipVKIE^*}$  represents the results obtained by using ground truth input instead of predicted input.
Our observations show that the performance of $\rm{PipVKIE^*}$  with ground truth input is better than that with predicted input, indicating that errors accumulate in downstream tasks. Furthermore, UniVKIE achieves better results than PipVKIE, demonstrating that unifying is a better strategy.

\subsubsection{EL}
The performance of EL is shown in the bottom right of Table \ref{tab:Main Results}. Similar to ER, we compared the performance of PipVKIE and UniVKIE when feeding them with either the ground truth entity boundaries or predicted hidden representations. Our observations show that the performance is slightly lower when using the predictions of ER. In real-world applications where errors can accumulate,  UniVKIE achieves better results than PipVKIE, which demonstrates its superiority.

\begin{table}
\centering
\scalebox{0.9}
{
    \begin{tabular}{cc|ccc}    
    \hline
    \multicolumn{2}{c|}{\textbf{Modals}} & \textbf{BTC} & \textbf{ER} & \textbf{EL}\\
    \hline
    {Visual} & {Text} & Acc & F1 & Acc \\
    \hline
    {\checkmark} & & 65.99 & - & - \\
    {} & \checkmark & 95.30 & 90.42 & 61.59 \\
    \checkmark & \checkmark & 97.22 & 94.58 & 71.61 \\
    \hline
    \end{tabular}
}
\caption{\label{tab: modal ablation study}
Modality ablation study of UniVKIE. - indicates the meaningless results, as the text modal cannot be omitted in ER and EL.
}
\end{table}

In Table \ref{tab:Main Results} UniVKIE outperforms PipVKIE in major metrics. We identified that this is primarily due to the efficient fusion of different modalities and the elimination of error accumulation caused by the pipeline method. Another factor is that the subtasks within PipVKIE operate independently and could not benefit from each other. 





\subsubsection{Ablation Study}
We design a series of ablation experiments to verify the contributions of each component in our solutions. We evaluate the effectiveness of modalities by eliminating one or some of them in UniVKIE, as illustrated in Table \ref{tab: modal ablation study}.
While text modal is necessary for ER and EL, we notice a manifest performance degradation in BTC after removing textual information, this confirms that the text modality plays a dominant role in our task. 
In addition, UniVKIE with multimodal information achieves the best results in all comparisons. To explore the reason, even for identical text, the visual features such as its location and background in a frame can affect the identification of segment categories, entities, and relationships. For example, subtitles are often located at the bottom of the image and have a special background color. Similarly, related names and identities often appear in visually adjacent positions within a frame of video.

\begin{table}
\centering
\scalebox{0.9}
{
    \begin{tabular}{ccc|ccc}
    \hline
     \multicolumn{3}{c|}{\textbf{Loss}} &  \textbf{BTC} & \textbf{ER} & \textbf{EL}\\
    \hline
    $\mathcal{L}_{BTC}$ & $\mathcal{L}_{ER}$ & $\mathcal{L}_{EL}$ & Acc & F1 & Acc\\
    \hline
    \checkmark & \checkmark & \checkmark & 97.22 & 94.58 & 71.61 \\

    & \checkmark & \checkmark & - & 93.46 & 73.29 \\

    \checkmark& \checkmark& & 97.05 & 94.19 & - \\

    \checkmark& & & 97.84& - & - \\
    \hline
    \end{tabular}
}
\caption{\label{tab: loss ablation study}
Loss ablation study of UniVKIE. - indicates the meaningless results as the task-specific loss is necessary for the corresponding subtask.
}
\end{table}
Furthermore, we conduct additional experiments to explore how each task impacts the others, which is shown in Table \ref{tab: loss ablation study}.  To explore the impact of BTC on ER and EL, we find that UniVKIE without BTC loss achieves slightly worse results on ER, but obtains improvement on EL. Moreover, by removing the ER loss and the EL loss, we find that the performance is almost steady on BTC. These phenomena indicate that BTC is hardly influenced by the other two tasks.  UniVKIE unifies the three tasks into one model and achieves overall balanced performance. 

\begin{table}[h]
\centering
\scalebox{0.9}
{
    \begin{tabular}{c|c|c}
    \hline
   \textbf{Methods} & \textbf{Speed} & \textbf{Params}\\
   \hline
   PipVKIE (BTC + ER + EL) & 205ms & 350M \\
   \hline
    UniVKIE (BTC + ER + EL) & 56ms & 106M \\
    \hline
    \end{tabular}
}
\caption{\label{citation-guide}
Efficiency comparison of PipVKIE and UniVKIE.
}
\end{table}
\section{Discussion}

\subsection{Modality} 
In the section of the ablation study, we find text modality plays the leading role. Besides, visual information also plays a crucial role in our task. For example, in BTC, box of a specific category often has a particular background color and location, which can serve as complementary features to the text. As the associated names and identities are usually located in associated position in one frame, it is important to consider visual information when performing EL tasks. The experimental results in Table \ref{tab: modal ablation study} validate this point.

\subsection{Efficiency} 
Table \ref{citation-guide} compares the inference speed and resource cost between PipVKIE and UniVKIE. We deploy both models on Tesla V100-SXM2-32GB. By sharing the same multimodal backbone and unifying the three tasks into one model,  UniVKIE achieves satisfactory inference speed and costs lower GPU resources. This is mainly attributed to the fact that in the inference of PipVKIE, only the feature in a single box is required, while in UniVKIE, the features of all boxes in a whole frame are inputted, which increases parallelism and thus improves efficiency. 

\subsection{Deployment Cases}
\begin{figure}[h]
        \centering
        \subfigure{
            \begin{minipage}[b]{7.5cm}
            \includegraphics[width=7.5cm]{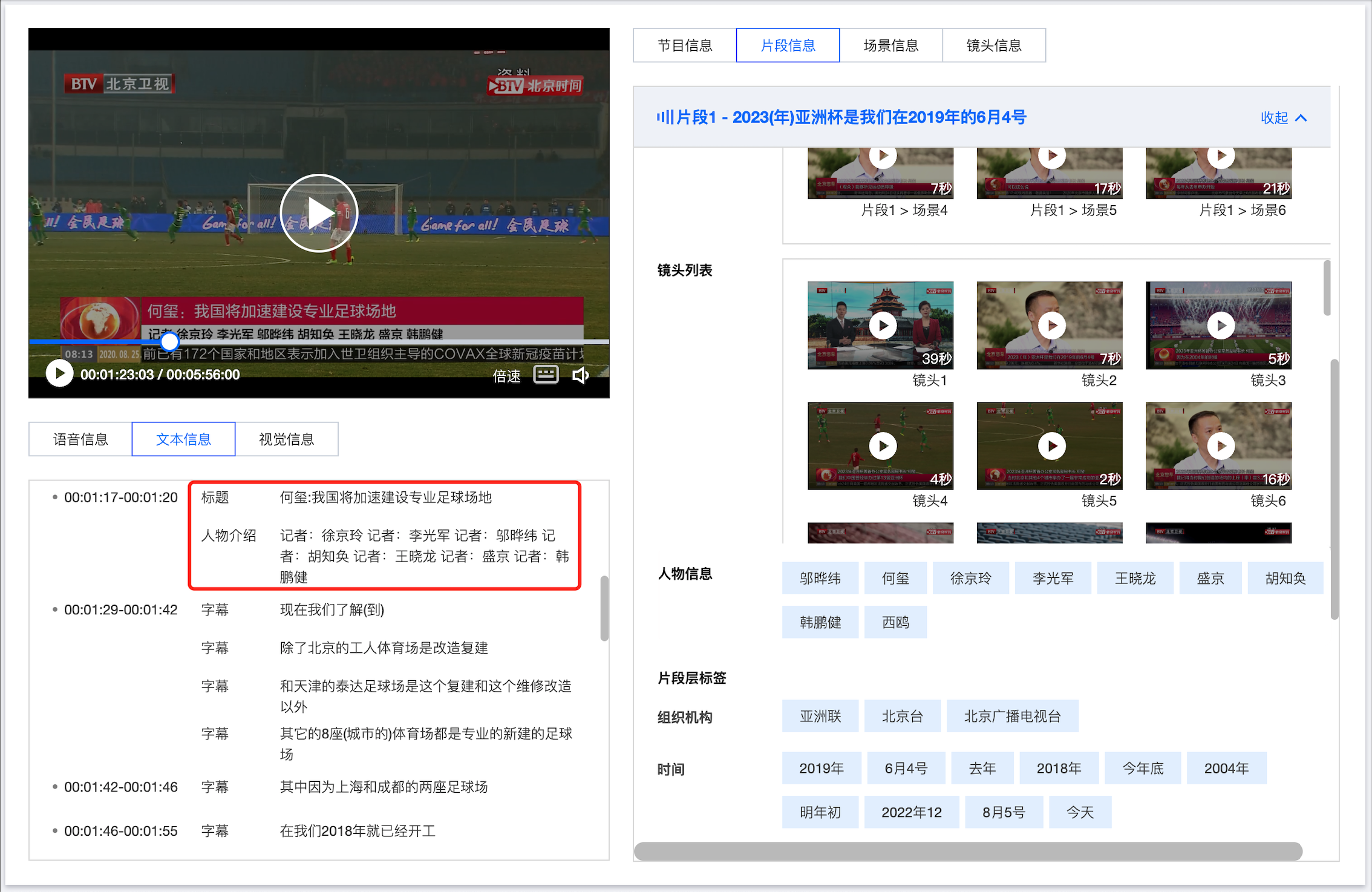} \\
            \vspace{-7mm}
            \end{minipage}
        }
        \subfigure{
            \begin{minipage}[b]{7.5cm}
            \includegraphics[width=7.5cm]{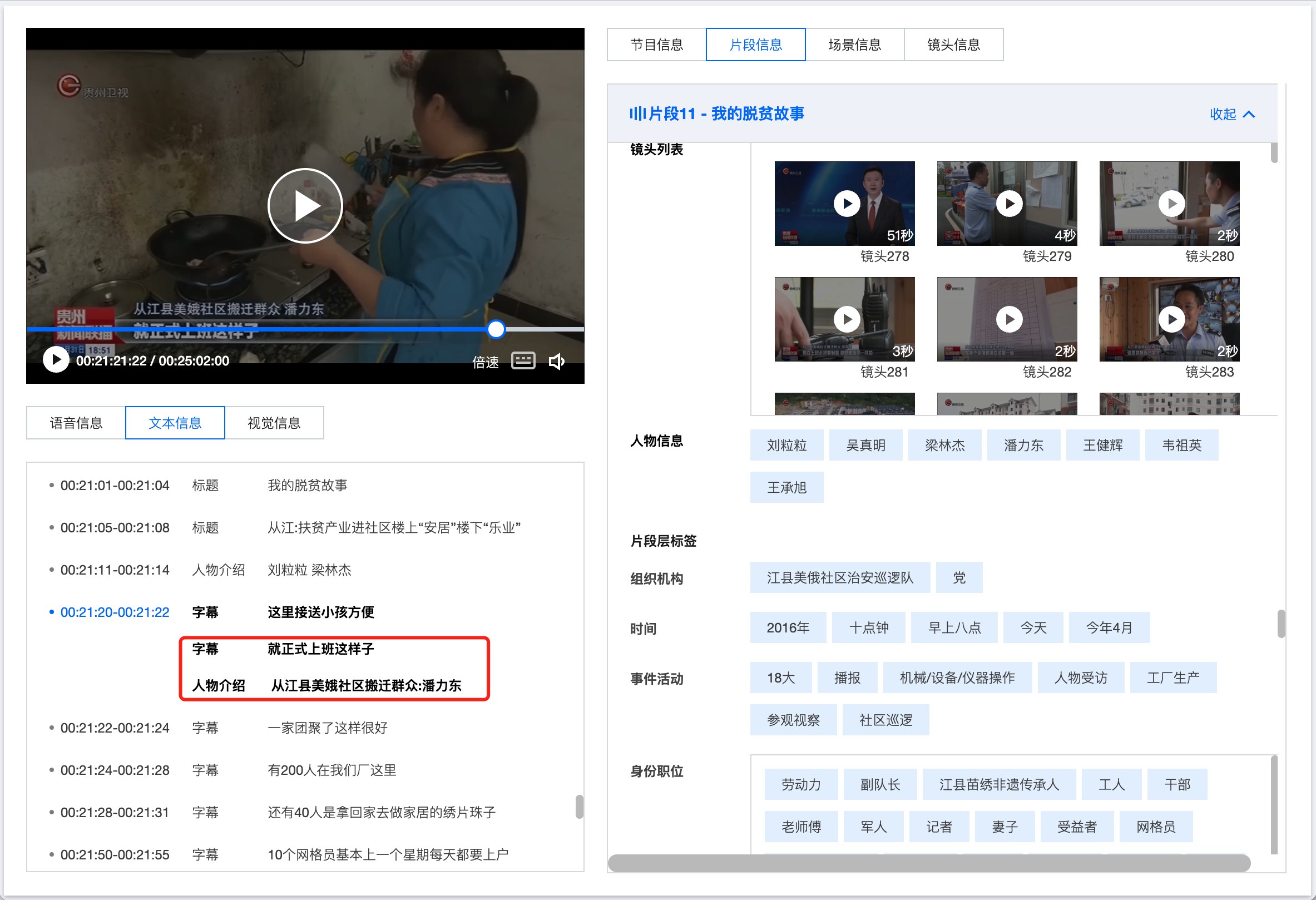} \\
            \vspace{-7mm}
            \end{minipage}
            }
        \vspace{-2ex}
        \subfigure{
            \begin{minipage}[b]{7.5cm}
            \includegraphics[width=7.5cm]{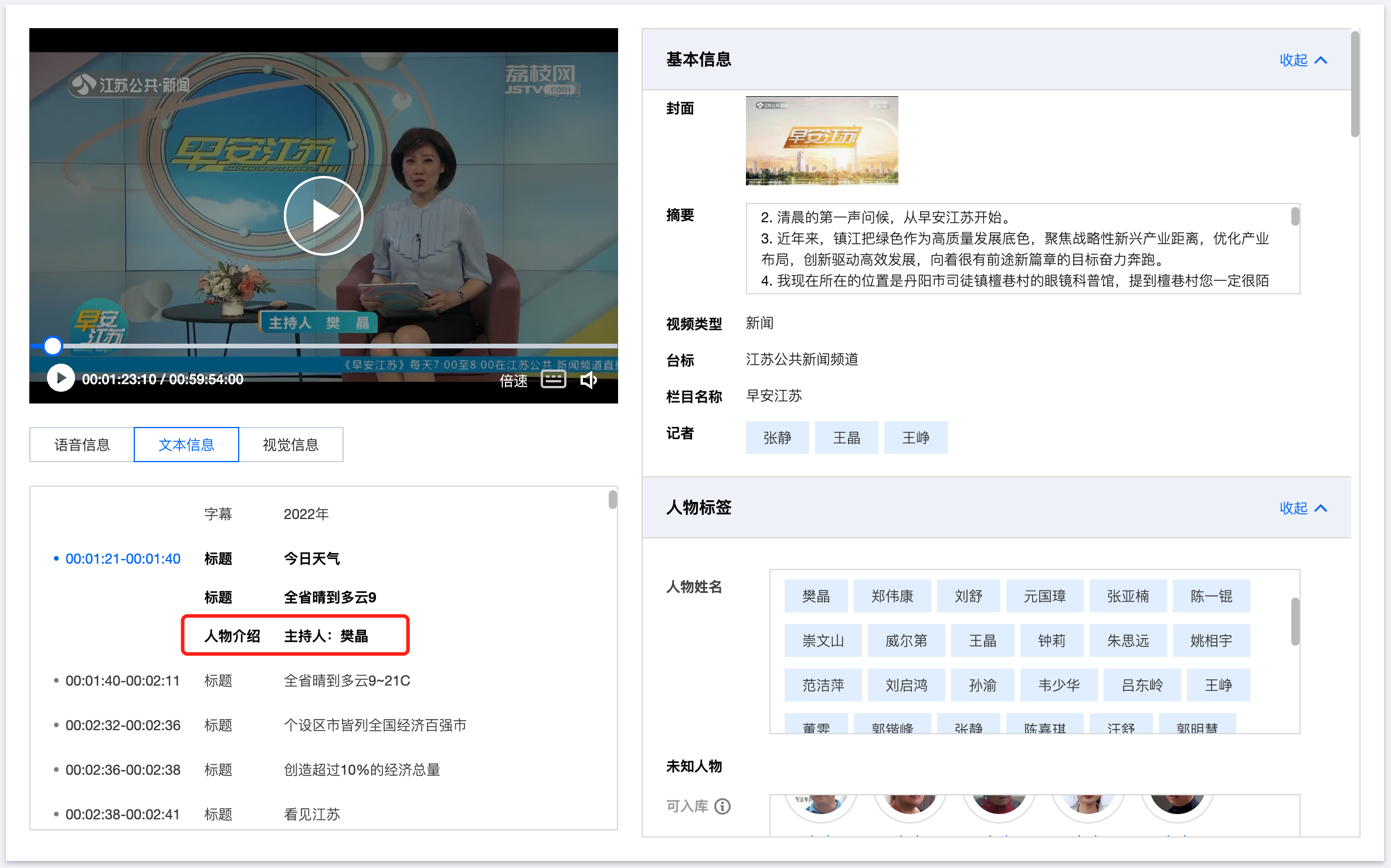} \\
            \vspace{-3mm}
            \end{minipage}
            }
        \caption{Real-world cases on our AI platform, the red boxes illustrate the extraction results of current frame.       \label{fig:cases on platform}}
\end{figure}

Both PipVKIE and UniVKIE have already been deployed on an AI platform for industrial media, which is a well-designed video understanding platform with comprehensive video processing services. 
We give three cases of real-world news videos, as shown in Fig.\ref{fig:cases on platform} . The red boxes illustrate the hierarchical information extracted from the current video frame. In these cases, titles and subtitles are shown at the segment level while personal information is organized at the entity level with name and identity. 
Therefore,  these valuable hierarchical information extracted by VKIE from the visual texts can be used effectively to index, organize, and search videos in real applications. More details about how our application works on the AI platform could be found in the supplementary material \ref{sec:platform}. 


\section{Conclusion}
This paper introduces a novel task in the industry, referred to as VKIE, which aims to extract crucial information from visual texts in videos. To address the task, we decouple VKIE into four subtasks: text detection and recognition, text classification, entity recognition, and relation extraction. Furthermore, we propose two complete solutions utilizing multimodal information: PipVKIE and UniVKIE. PipVKIE  performs these three subtasks in different stages, while UniVKIE unifies all of them in one model with higher efficiency and lower resource cost. 
Experimental results on one well-defined dataset demonstrate that our solutions can achieve remarkable performance and satisfactory resource cost. With VKIE, structured information could be effectively extracted and well organized with rich semantic information. VKIE has been deployed on an industrial AI platform.

\section*{Limitations}
While VKIE could be easily extended to multilingual tasks, our dataset in practical application centered on Chinese videos. For general use, we are formulating plans to extend the application to multilingual tasks in the future. 



\section*{Ethics Statement}
The authors declare that the data in our work is publicly available and does not involve political and moral sensitivities. 
Ethical concerns include the usage of the proposed solution for a purpose different from that previously mentioned in the paper, such as video inputs of racism, violence, etc.


\appendix


\section{Media AI Platform}
\label{sec:platform}
\begin{figure*}[h]
    \centering
    \includegraphics[width=1\linewidth, frame]{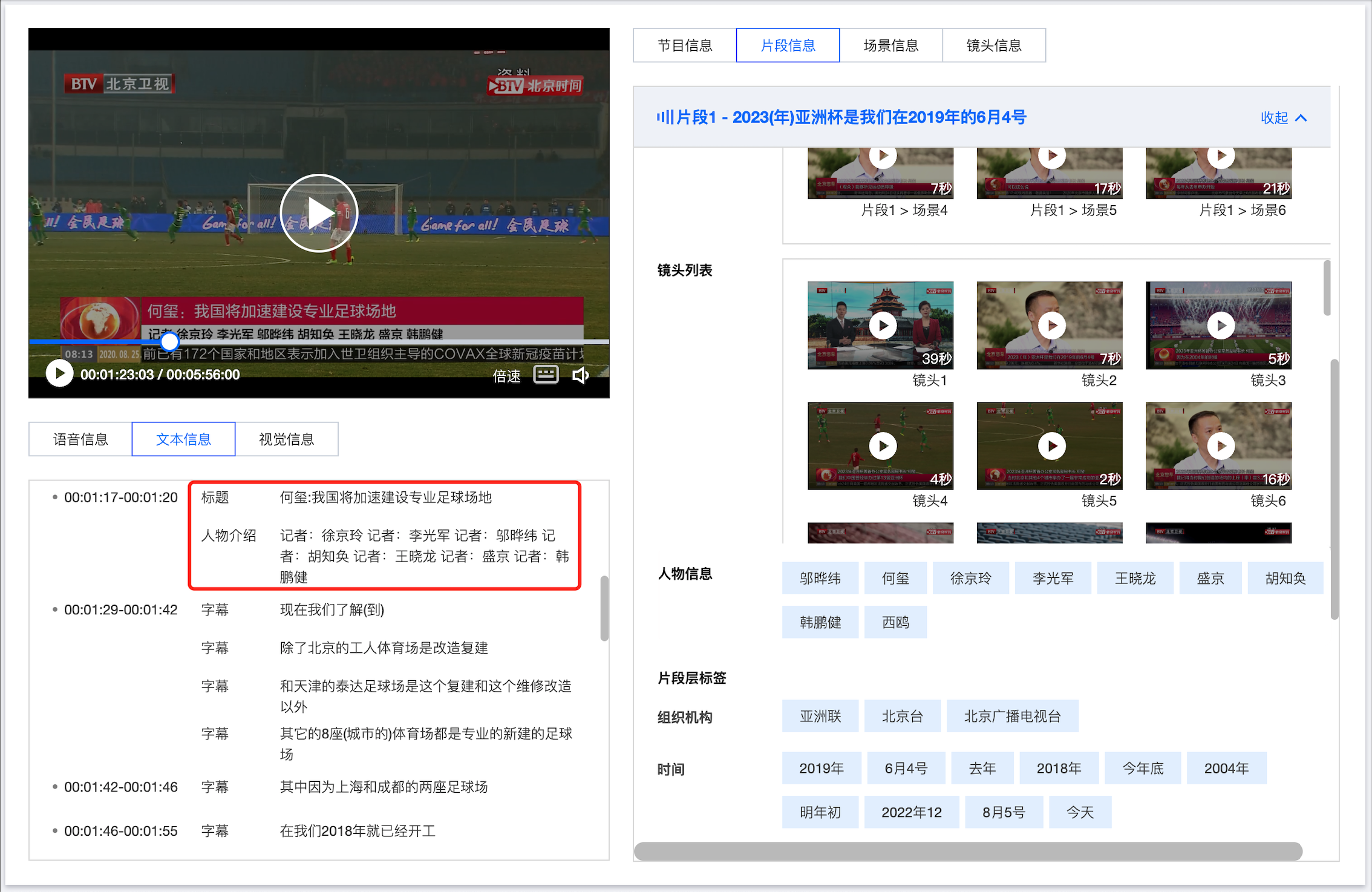}
    \centering
    \caption{A real-world case on the AI platform for clear viewing.
    \label{fig:complete cases on platform}}
\end{figure*}

The task of key information extraction from  visual texts in videos has been  deployed on an media AI platform, which is a well-designed video understanding platform with comprehensive video processing services.  We uniformly sample key frames from the uploaded video. Then, a OCR engine is used to extract visual boxes and their corresponding coordinates. Afterwards, VKIE completes the three subtasks of BTC, ER and EL, and obtains  hierarchical information at the entity and segment levels. Here we present one result for clear viewing as in Fig.\ref{fig:complete cases on platform}.

\section{Details of dataset}
\label{sec:dataset details}

\begin{figure*}[h]
    \centering
    \includegraphics[width=1\linewidth, frame]{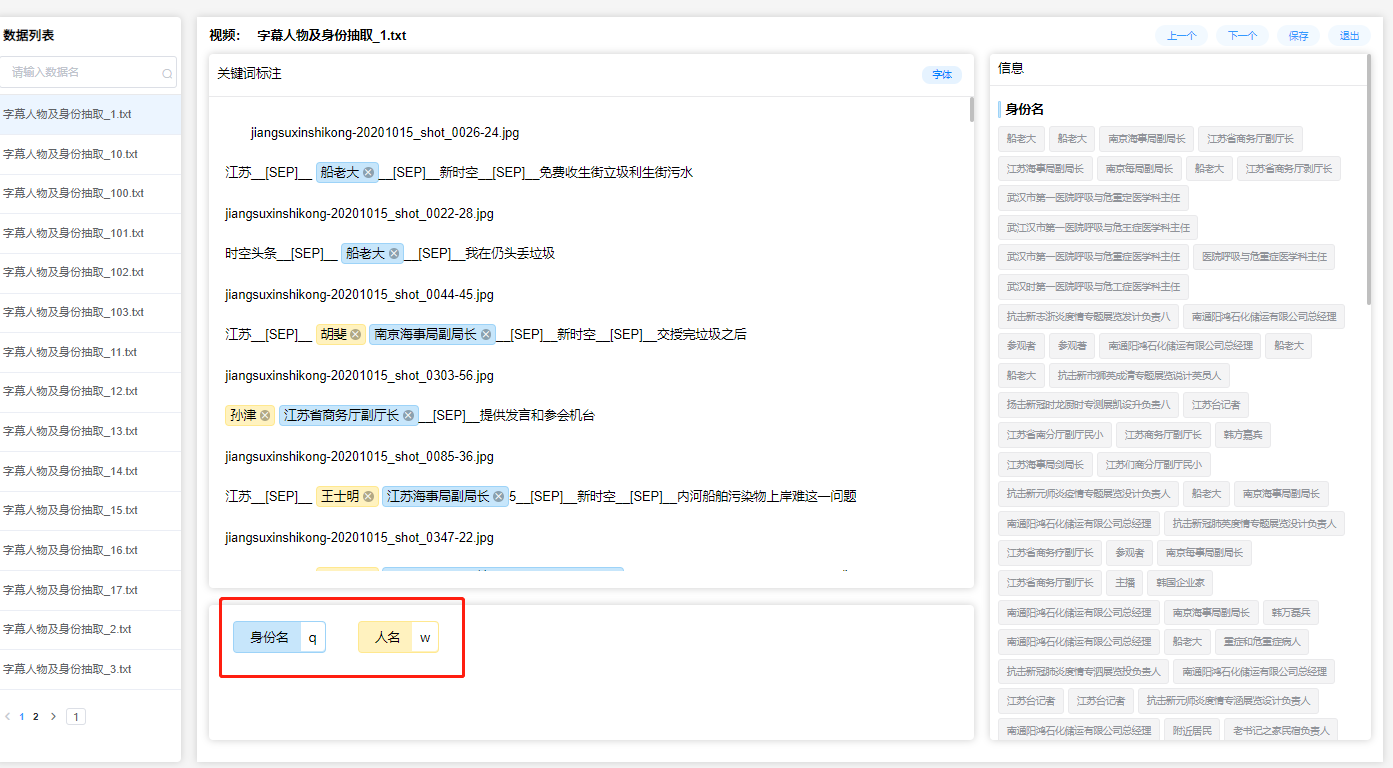}
    \centering
    \caption{Examples of ER on the annotation platform. The red box indicates the candidate labels of ER.
    \label{annotation_case2}}
\end{figure*}

Table \ref{tab:dataset description} illustrates the concrete categories contained in \textbf{BTC}, \textbf{ER}, and \textbf{EL} in our practice. We collected 88 sources, totaling 115 hours, from publicly available videos, including news programs, variety shows, and other sources.
All 88 video sources are split for training, developing, and testing with the ratio 3:1:1. We then extract frames from these videos by taking their average over time. To prevent data leakage, we ensure that frames from the same video are not present in different splits.
In \textbf{BTC}, we assign the samples to 4 categories including Title, Person Info, Subtitle, and Misc.  We further annotate mentions and labels on the samples of Person Info for  \textbf{ER} as shown in Fig.\ref{annotation_case2} . Finally,  \textbf{EL} is annotated on the pairs of entities extracted from each frame.

\begin{table}[h]
\scalebox{0.85}
{
\begin{tabular}{lll}
\toprule
Task       & Type          & Value                                          \\
       \midrule
Video  & Total Hours   & 115                                            \\
       & Total Sources & 88                                             \\
       & Total Videos  & 264 \\
       & Total Frames  & 23896   \\
       \midrule
BTC    & Categories    & Title, Personal Info\\
    &     & Subtitle, Misc \\
       & Samples       & train/dev/test: 76k/22k/25k                    \\
       \midrule
ER & Categories        & Name, Identity                              \\
       & Samples       & train/dev/test: 34k/12k/12k                     \\
       \midrule
EL     & Categories    & Matched, Not matched                           \\
       & Samples       & train/dev/test: 18k/6k/6k                      \\
\bottomrule
\end{tabular}
}
\caption{\label{tab:dataset description}
The basic statistics of our datasets
}
\label{tab:dataset description}
\end{table}


\section{Training Hyperparameters}
Table \ref{tab:hyperparameters of PipVKIE} illustrates the hyperparameters of the three models corresponding to  \textbf{BTC}, \textbf{ER} and \textbf{EL} in PipVKIE.  In UniVKIE, we use a shared multimodal backbone and build task-specific branches as in Table \ref{tab:hyperparameters of UniVKIE}.

\begin{table}[h]
\small\centering
  \label{Hyperparameters of PipV2TKIE}
  \resizebox{0.5\textwidth}{!}{
  \begin{tabular}{ll}
  \toprule
Hyperparameters &Value\\
    \midrule
    \textbf{BTC} & \\
    visual feature extractor& 3-layers CNN  \\
    textual feature extractor & 4-layers transformers \\
    hidden dimension of visual feature & 266 \\
    hidden dimension of textual feature & 768 \\
    optimizer & Adam \\
    batch size & 48\\
    epochs of training & 10 \\
    \textbf{ER} & \\
    textual feature extractor & transformer \\  
    hidden dimension of textual feature & 768 \\
    optimizer & AdamW \\
    batch size & 16\\
    epochs of training & 10 \\
    \textbf{EL} & \\
    textual feature extractor & transformer \\  
    hidden dimension of textual feature & 768 \\
    optimizer & AdamW \\
    batch size & 16 \\
    epochs of training & 10 \\
  \bottomrule
\end{tabular}}

\caption{Hyperparameters of PipVKIE}
\label{tab:hyperparameters of PipVKIE}
\end{table}

\begin{table}[h]
\small\centering
  \label{Hyperparameters of UniVKIE}
  \resizebox{0.5\textwidth}{!}{
  \begin{tabular}{ll}
    \toprule
Hyperparameters &Value\\
    \midrule
    image channels & 3 \\
    normalized coordinate size & 128 \\  
    hidden dimension of multimodal feature & 768 \\
    batch size & 32 \\
    epochs of training & 10 \\
    optimizer & AdamW \\
    learning rate & 5e-5 \\
    hidden layer dropout prob & 0.1 \\
    number of hidden layers & 12 \\
    hidden dimension & 768 \\
    token max length in encoder & 128 \\
    2d position embedding dimension & 1024 \\
    1d position embedding dimension & 512 \\
    vocabulary size & 21128 \\
    BTC/ER/EL trade-off factors in loss & 0.3/0.3/0.4 \\
    \bottomrule
\end{tabular}}
\caption{Hyperparameters of UniVKIE}
\label{tab:hyperparameters of UniVKIE}
\end{table}

\section{Integration with LLMs}
Recently, large language models(LLMs) have attracted widespread interest. We have noticed this and conducted experiments with LLMs within the VKIE scenario. 
However, we found these approaches are not sufficiently stable for practical industrial applications. Therefore, we have decided to defer the exploration of integration with LLMs as a future extension of our work, rather than incorporating it into this submission.

\end{document}